\documentclass{PoS}

\usepackage{graphics,epstopdf,color,cancel,graphicx}
\usepackage[utf8]{inputenc}
\usepackage{lmodern}
\usepackage{textcomp}
\usepackage{float}
\usepackage{lineno}
\usepackage{siunitx}
\usepackage{aas}
\usepackage[numbers]{natbib}
\newcommand{\be}{\begin{equation}}
\newcommand{\ee}{\end{equation}}
\newcommand{\bary}{\begin{eqnarray}}
\newcommand{\eary}{\end{eqnarray}}

\title{Monitoring radio galaxies at TeV energies with HAWC}

\ShortTitle{Monitoring radio}

\author{\speaker{Daniel Avila Rojas}, Rubén Alfaro\\
        Instituto de Física, UNAM, Ciudad de México, México.\\
        E-mail: \email{daniel\_avila5@ciencias.unam.mx, ruben@fisica.unam.mx}}

\author{M. Magdalena González, Nissim Fraija\\
        Instituto de Astronomía, UNAM, Ciudad de México, México.\\
        E-mail: \email{magda@astro.unam.mx, nifraija@astro.unam.mx}}

\author{for the HAWC Collaboration \thanks{For a complete author list, see http://www.hawc-observatory.org/collaboration/icrc2019.php}\\
        }

\abstract{With an instantaneous field of view of ~2 sr and a duty cycle > 95\%, the High Altitude Water Cherenkov (HAWC) Gamma-Ray Observatory is a perfect instrument for monitoring variable TeV sources. Because radio galaxies are a type of Active Galactic Nuclei (AGN) with their jets misaligned with respect to our line of sight, they may help us to probe the physics of very-high-energy (VHE) emission processes in these objects. Three out of four radio galaxies that have been detected at TeV energies by other facilities are located within the field of view of the HAWC Observatory: M87, NGC 1275, and 3C 264. A search for TeV gamma rays at their locations yields no statistically significant excess of counts. We present corresponding upper limits for each radio galaxy and light curves covering 3 years of data taken with HAWC.}

\FullConference{36th International Cosmic Ray Conference -ICRC2019-\\
		July 24th - August 1st, 2019\\
		Madison, WI, U.S.A.}

\begin{document}

\section{Introduction}
According to unification models, radio galaxies are a type of Active Galactic Nuclei (AGNs) presenting radio-loud emission and with its jet misaligned with the line of sight. Morphologically, they are classified according to their radio emission into two categories. FR-I with the bright radio emission dominating close to its center and FR-II with the radio emission peak further away at the lobes\cite{1974MNRAS.167P..31F}. Although they were not expected to emit in TeV energies, recently four FR-I galaxies have been detected at these energies. Thus, an interesting debate concerning the very-high-energy (VHE) gamma-rays emission models has been opened. This emission is believed to be mainly originated in the jets and lobes. The spectral energy distribution (SED) of radio galaxies in this energetic band can be well-described by leptonic models, hadronic models or a combination of both. Leptonic models can explain gamma-ray emission up to GeV energy range by means of synchrotron self-Compton (SSC) emission \cite{2014MNRAS.441.1209F, 2017APh....89...14F}, while hadronic models describe emission at TeV energies as originated by photo-hadronic processes \cite{2016ApJ...830...81F, 2014A&A...562A..12P}.

The HAWC Observatory, with its instantaneous field of view of $\sim$ 2 sr and a duty cycle $\gtrsim$ 95\%, is well-suited for monitoring this kind of sources. In the present work we search for TeV emission from radio galaxies within the field of view of HAWC. Because no statistically significant excess of counts was collected, we report the upper limits on the VHE flux, as well as the light curves, for the radio galaxies M87, NGC 1275, and 3C 264 using 1017 days of data. The outline is as follows. In Section 2, we give a brief summary of the latest TeV observations for each radio galaxy as well as the explanation of the analysis, and in Section 3 we summarize our results.

\section{Radio Galaxies Observations and Analysis}

\subsection{M87}

M87 is a giant elliptical radio galaxy located in the Virgo cluster that harbors in its core a supermassive black hole (SMBH) with a mass of $ (6.5 \pm 0.7) \times 10^{9} \; \si{M}_{\odot} $ \cite {2019ApJ...875L...1E}. Its distance to Earth is $ 16.7 \pm 0.2 \; \si{Mpc} $ with a redshift of $ z = 0.0044 $ \cite {2007ApJ...655..144M}, makes it the closest radio galaxy in the field of view of HAWC. A relativistic jet emerges from its core and extends from $ 1.5$ to $ 2 \; \si{kpc} $ at an angle that has been estimated between $ \ang{15} - \ang{25} $ from the line of sight.  

HEGRA first detected M87 above 730 $ \si{GeV} $ reporting $3.3\% \pm 0.8\%$ of the Crab Nebula flux. Since then, it has been monitored by different imaging atmospheric Cherenkov telescopes (IACTs), quiescent and active states have been observed during these campaigns. Because of its closeness to Earth this radio galaxy is of great interest, affording an excellent opportunity to study the mechanisms involved on the emission of VHE gamma-rays. 

For a quiescent state M87 spectra have been fitted to a simple power law. In 2004 HESS reported the lowest flux ever registered with a spectral index of $ 2.62 \pm 0.35$ and a normalization of $(2.43 \pm 0.75) \times 10^{-13} \; \si{cm^{-2}.s^{-1}.TeV^{-1}} $ at $1$ $\si{TeV}$ \citep{2005tsra.conf..650B}. MAGIC monitored M87 between 2005 and 2007 and reported a spectral index of $ 2.21 \pm 0.21$ and a normalization of $(5.4 \pm 1.1) \times 10^{-12} \; \si{cm^{-2}.s^{-1}.TeV^{-1}} $ at $1$ $\si{TeV}$ \citep{2012A&A...544A..96A}. In 2007 VERITAS monitored M87 and reported a spectral index of $ 2.31 \pm 0.17$ and a normalization of $(7.4 \pm 1.3) \times 10^{-13} \; \si{cm^{-2}.s^{-1}.TeV^{-1}} $ at $1$ $\si{TeV}$ \citep{2008ApJ...679..397A}. Between 2011 and 2012, M87 was monitored by VERITAS, monthly variation in the TeV flux was observed which may hint that the quiescent emission evolves over longer time scales in comparison with flare emission. VERITAS divided their data sets in two from the 2012 observations; the first one with a spectral index of $ 2.1 \pm 0.3$ and a normalization of $(6.3 \pm 1.6) \times 10^{-13} \; \si{cm^{-2}.s^{-1}.TeV^{-1}} $ at $1$ $\si{TeV}$, and the second one with a spectral index of $ 2.6 \pm 0.2$ and a normalization of $(7.0 \pm 1.5) \times 10^{-13} \; \si{cm^{-2}.s^{-1}.TeV^{-1}} $ at $1$ $\si{TeV}$ \citep{2012AIPC.1505..586B}.

M87 was in activity or flare states in 2005, 2008 and 2010. The spectra for these flares were fitted to simple-power laws. For the 2005 flare, HESS reported a spectral index of $ 2.22 \pm 0.15$ and a normalization of $(11.7 \pm 1.6) \times 10^{-13} \; \si{cm^{-2}.s^{-1}.TeV^{-1}} $ at $1$ $\si{TeV}$ \citep{2006Sci...314.1424A}. For the 2008 flare MAGIC, reported a spectral index of $ 2.21 \pm 0.18$ and a normalization of $(48.1 \pm 8.2) \times 10^{-13} \; \si{cm^{-2}.s^{-1}.TeV^{-1}} $ at $1$ $\si{TeV}$ \citep{2008ApJ...685L..23A}, while VERITAS reported a spectral index of $ 2.40 \pm 0.21$ and a normalization of $(15.9 \pm 2.9) \times 10^{-13} \; \si{cm^{-2}.s^{-1}.TeV^{-1}} $ at $1$ $\si{TeV}$ \citep{2010ApJ...716..819A}. For the 2010 flare, VERITAS reported a spectral index of $ 2.19 \pm 0.07$ and a normalization of $(47.1 \pm 2.9) \times 10^{-13} \; \si{cm^{-2}.s^{-1}.TeV^{-1}} $ at $1$ $\si{TeV}$ \citep{2012ApJ...746..141A}. 

\subsection{NGC 1275}

NGC 1275 is located in the center of the Perseus cluster with a redshift of $ z = 0.017559 $, at a distance of $ \sim 75.3 \; \si{Mpc} $ \cite{1992ApJS...83...29S}. Its equatorial coordinates are $ \alpha = \ang{49,950} $ and $ \delta = \ang{41.512} $ (J2000). The mass of the SMBH in its core is $ 3-4 \times 10^{8} \; \si{M}_{\odot} $ \cite{2005MNRAS.359..755W} and the parsec scale jet is oriented with an angle of $ \sim \ang{30} - \ang{60} $ with respect to the line of sight. 
It was first detected by MAGIC above $ 100 \; \si{GeV} $ with a significance of $ 6.6 \;\sigma $ with nearly $ 100 \; \si{h} $ of data. Its flux between 70 and 500 $ \si{GeV}$ could be described by a simple power law  with a spectral index of $ 4.1 \pm 0.7_{\rm stat} \pm 0.3_{\rm sys} $ and a normalization of $(3.1 \pm 1.0_{\rm stat} \pm 0.7_{\rm sys}) \times 10^{-10} \; \si{TeV^{-1}\, cm^{-2}\,s^{-1}} $ at $ 100 \; \si{GeV} $\cite{2012A&A...539L...2A}.

Two flare activity periods have been reported: the first one in October 2016, was reported with $16\%$ of the Crab Nebula flux and the second one in the night of December $31^{st}$, 2016, with $\sim 1.5$ of the Crab Nebula flux.  The last flare correspond to the highest state ever reported for this source. The MAGIC Collaboration reported a spectrum fitted by a power law with an exponential cutoff with a spectral index of $ 2.11 \pm 0.14 $, a normalization of $(1.61 \pm 0.23) \times 10^{-9} \; \si{cm^{-2}.s^{-1}.TeV^{-1}} $ at $ 300 \; \si{GeV} $ and a cutoff energy of $ 0.56 \pm 0.11 \; \si{TeV}$. A similar analysis was carried out including Fermi-LAT data, the spectrum was fitted with  a power-law function with exponential cutoff, a spectral power index of $ 2.05 \pm 0.03 $, a normalization of $(4.17 \pm 0.22) \times 10^{-9} \; \si{cm^{-2}.s^{-1}.TeV^{-1}} $ at $ 198.21 \; \si{GeV} $ and a cutoff energy of $ 492 \pm 35 \; \si{GeV} $ \cite{2018A&A...617A..91M}.

\subsection{3C 264}

3C 264 is a FR-I radio galaxy located in the Leo cluster at a redshift of $ z = 0.022 $ corresponding to a distance of $ \sim 95 \; \si{Mpc} $ \cite{1999ApJS..125...35S}. This makes it the most distant radio galaxy detected so far at VHE although the true nature of PKS 0625 35 is still under debate. Its equatorial coordinates are $ \alpha = \ang{176.271} $ and $ \delta = \ang{19.606} $ (J2000). It harbors a SMBH with a mass of $ \sim 5 \times 10^{8} \; \si{M}_{\odot} $ \cite{2015A&A...581A..33D}. It has a relativistic jet that reaches kiloparsec scales, in which knots have been observed and those closest to the core present superluminal movement \cite{2015Natur.521..495M}. 

The only instrument that has observed this source at TeV energies is VERITAS, which observed it for a period of time of $ \sim 12 \; \si{h} $ between February and March 2018, making it the most recent galaxy radio to join the TeV emitters. The preliminary results showed an excess of 60 gamma-ray events on the background, with a significance of $5.4 \sigma $. The preliminary integral flux is $ \sim 1 \% $ of the Crab Nebula flux \cite{2018ATel11436....1M}. It has been argued that the VHE spectral index is $\sim 2.3$ and shows a low, weakly variable flux along with some month-scale variations \cite{2018arXiv181005409R}. 

\subsection{Analysis and Results}

We performed a search for TeV gamma-rays from the radio galaxies M87, NGC 1275 and 3C 264 using 1017 days of data via a maximum likelihood fit convolving spectral models applied to the data with the detector response of HAWC as described in \cite{2017ApJ...843...39A}. No signal with enough statistical significance from these radio galaxies were detected, thus, 95\% confidence level upper limits were calculated using a maximum likelihood method within HAWC analysis framework \citep{2015arXiv150807479Y}. For the light curves we use the same procedure as the one used in \citep{2017ApJ...841..100A}.  We also applied a quality cut to the data for which the sources transit for at least a sidereal fraction of $0.75$. In accordance to previous works \citep{2008ApJ...679..397A, 2018arXiv181005409R, 2018A&A...617A..91M}, the spectral models used for the analysis are: i) for M87, a simple power law with spectral index $\Gamma =2.31$, ii) for NGC 1275, a power law with exponential cutoff at $500 \; \si{GeV}$ and spectral index $\Gamma =3$ and iii) for 3C 264, a simple power law with spectral index $\Gamma = 2.3$. 

We summarize on Table \ref{upli} the resulting VHE upper limits on the flux normalization for the radio galaxies M87, NGC 1275 and 3C 264. The VHE upper limits were computed using 1017 days of data taken with the HAWC Observatory. Given that these radio galaxies are potential candidates for accelerating cosmic rays, and then for producing VHE photons by hadronic interactions, an estimate of the characteristics of the emitting region, magnetic fields and the amount of cosmic rays in the jet could be given.

\begin{table}[H]
\centering
\caption{Upper limits calculated for the radio galaxies within the field of view of HAWC. For all radio galaxies the attenuation in the flux due to interaction with the extragalactic background light (EBL) was considered. For M87 the first value of the upper limit considers no EBL attenuation.}\label{upli}
\begin{tabular}{lccc}
\hline\hline
Radio Galaxy & Energy Range & Spectral Index & Upper Limit\\
& [TeV] & & [$10^{-13}\,{\rm TeV^{-1}\,cm^{-2}\,s^{-1}}$]\\
\hline
M87 & 3 - 100 & $2.31$ & $1.76$ \\ 
M87 (EBL) & 3 - 100 & $2.31$ & $3.51$\\
\hline
NGC 1275 (EBL) & 1 - 4 & $3$ & $167.8$\\
\hline
3C 264 (EBL) & 2 - 40 & $2.3$ & $6.88$\\
\hline
\end{tabular}
\end{table}

Energy range were chosen as the one that provides a 90\% of the sensitivity for a given spectral shape and declination of the source. The upper limit for M87 is consistent with past observations of the source in a quiescent state. For NGC 1275 the upper limit is comparable with the flux of the 2017 flare. The upper limit for 3C 264 will be compared with the data taken by VERITAS as soon as their results get published.

\begin{figure}[H]
  \centering
    \includegraphics[width=0.95\textwidth]%
                    {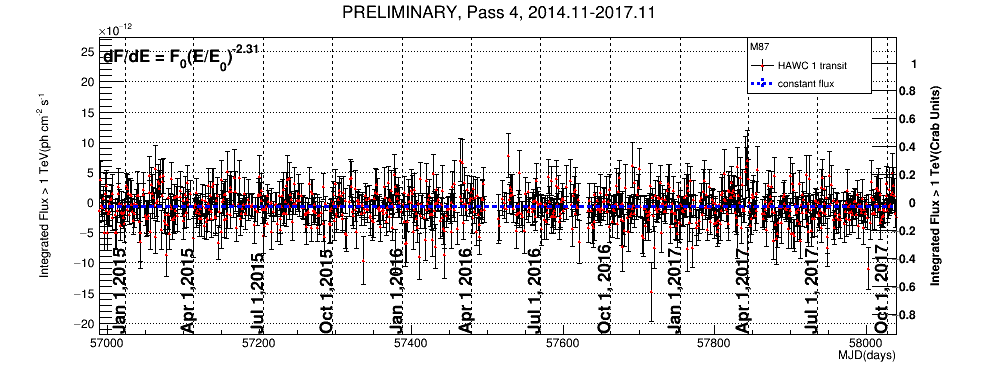}
\caption{Light curve for 1017 days of HAWC data for M87. \label{cap3:fig:M87LC1}}
\end{figure}

\begin{figure}[H]
  \centering
    \includegraphics[width=0.95\textwidth]%
                    {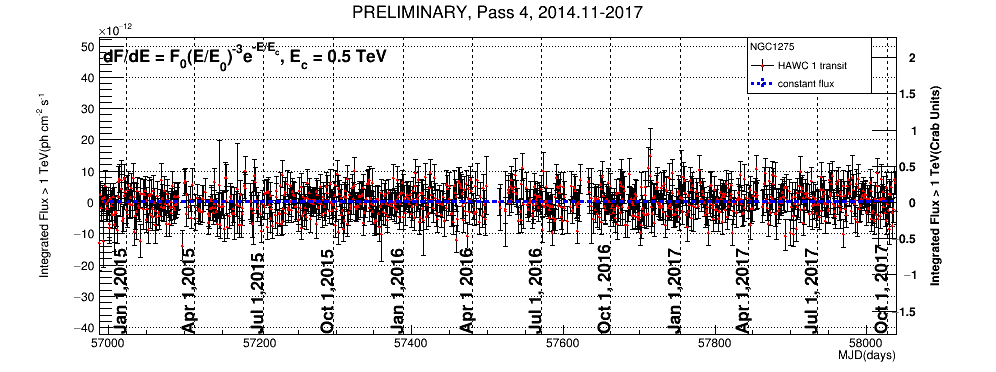}
\caption{Light curve for 1017 days of HAWC data for NGC 1275. \label{cap3:fig:NGCLC}}
\end{figure}

\begin{figure}[H]
  \centering
    \includegraphics[width=0.95\textwidth]%
                    {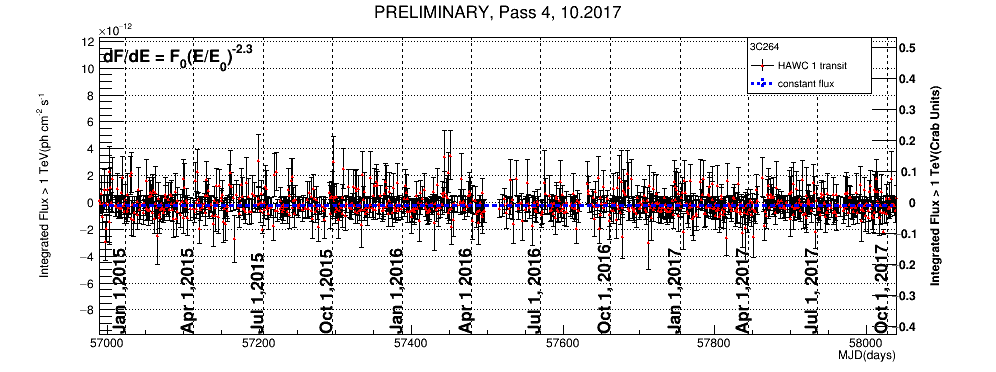}
\caption{Light curve for 1017 days of HAWC data for 3C 264. \label{cap3:fig:3CLC}}
\end{figure}

All of the light curves have mean flux values consistent with zero. There are no signatures of evident flare events, however, a Bayesian analysis will be done to dismiss the possibility of any hidden activity in the light curves.


\section{Conclusions}

The upper limits calculated in the GeV - TeV energy range would help to constrain the physics of these radio galaxies (e.g., the emission processes, the size of the emitting region, magnetic fields, the amount of cosmic rays, etc). The light curves derived in this work will be updated using the ZEBRA framework as done in \cite{JA_ICRC}. Weekly and monthly light curves will be done to search for variability of the sources in these time scales.

HAWC gamma-ray observatory will continue monitoring the radio galaxies in its field of view with the goal of detecting VHE photons in the following years.

\section{Acknowlegments}
We acknowledge the support from: the US National Science Foundation (NSF) the US Department of Energy Office of High-Energy Physics; 
the Laboratory Directed Research and Development (LDRD) program of Los Alamos National Laboratory; 
Consejo Nacional de Ciencia y Tecnolog\'{\i}a (CONACyT), M{\'e}xico (grants 271051, 232656, 260378, 179588, 239762, 254964, 271737, 258865, 243290, 132197, 281653)(C{\'a}tedras 873, 1563, 341), Laboratorio Nacional HAWC de rayos gamma; 
L'OREAL Fellowship for Women in Science 2014; 
Red HAWC, M{\'e}xico; 
DGAPA-UNAM (grants AG100317, IN111315, IN111716-3, IA102715, IN111419, IA102019, IN112218); 
VIEP-BUAP; 
PIFI 2012, 2013, PROFOCIE 2014, 2015; 
the University of Wisconsin Alumni Research Foundation; 
the Institute of Geophysics, Planetary Physics, and Signatures at Los Alamos National Laboratory; 
Polish Science Centre grant DEC-2014/13/B/ST9/945, DEC-2017/27/B/ST9/02272; 
Coordinaci{\'o}n de la Investigaci{\'o}n Cient\'{\i}fica de la Universidad Michoacana; Royal Society - Newton Advanced Fellowship 180385. Thanks to Scott Delay, Luciano D\'{\i}az and Eduardo Murrieta for technical support.


\bibliography{biblio}

\begin{thebibliography}{28}
\providecommand{\natexlab}[1]{#1}
\providecommand{\url}[1]{\texttt{#1}}
\expandafter\ifx\csname urlstyle\endcsname\relax
  \providecommand{\doi}[1]{doi: #1}\else
  \providecommand{\doi}{doi: \begingroup \urlstyle{rm}\Url}\fi

\bibitem[{Fanaroff} and {Riley}(1974)]{1974MNRAS.167P..31F}
B.~L. {Fanaroff} and J.~M. {Riley}.
\newblock {The morphology of extragalactic radio sources of high and low
  luminosity}.
\newblock \emph{\mnras}, 167:\penalty0 31P--36P, May 1974.
\newblock \doi{10.1093/mnras/167.1.31P}.

\bibitem[{Fraija}(2014)]{2014MNRAS.441.1209F}
N.~{Fraija}.
\newblock {Gamma-ray fluxes from the core emission of Centaurus A: a puzzle
  solved}.
\newblock \emph{\mnras}, 441:\penalty0 1209--1216, June 2014.
\newblock \doi{10.1093/mnras/stu652}.

\bibitem[{Fraija} et~al.(2017){Fraija}, {Marinelli}, {Galv{\'a}n-G{\'a}mez},
  and {Aguilar-Ruiz}]{2017APh....89...14F}
N.~{Fraija}, A.~{Marinelli}, A.~{Galv{\'a}n-G{\'a}mez}, and E.~{Aguilar-Ruiz}.
\newblock {Modeling the spectral energy distribution of the radio galaxy
  IC310}.
\newblock \emph{Astroparticle Physics}, 89:\penalty0 14--22, Mar 2017.
\newblock \doi{10.1016/j.astropartphys.2017.01.001}.

\bibitem[{Fraija} and {Marinelli}(2016)]{2016ApJ...830...81F}
N.~{Fraija} and A.~{Marinelli}.
\newblock {Neutrino, {\ensuremath{\gamma}}-Ray, and Cosmic-Ray Fluxes from the
  Core of the Closest Radio Galaxies}.
\newblock \emph{\apj}, 830\penalty0 (2):\penalty0 81, Oct 2016.
\newblock \doi{10.3847/0004-637X/830/2/81}.

\bibitem[{Petropoulou} et~al.(2014){Petropoulou}, {Lefa}, {Dimitrakoudis}, and
  {Mastichiadis}]{2014A&A...562A..12P}
M.~{Petropoulou}, E.~{Lefa}, S.~{Dimitrakoudis}, and A.~{Mastichiadis}.
\newblock {One-zone synchrotron self-Compton model for the core emission of
  Centaurus A revisited}.
\newblock \emph{\aap}, 562:\penalty0 A12, February 2014.
\newblock \doi{10.1051/0004-6361/201322833}.

\bibitem[{Event Horizon Telescope Collaboration} et~al.(2019){Event Horizon
  Telescope Collaboration}, {Akiyama}, {Alberdi}, {Alef}, {Asada}, {Azulay},
  {Baczko}, {Ball}, {Balokovi{\'c}}, {Barrett}, and
  et~al.]{2019ApJ...875L...1E}
{Event Horizon Telescope Collaboration}, K.~{Akiyama}, A.~{Alberdi}, et~al.
\newblock {First M87 Event Horizon Telescope Results. I. The Shadow of the
  Supermassive Black Hole}.
\newblock \emph{\apjl}, 875:\penalty0 L1, April 2019.
\newblock \doi{10.3847/2041-8213/ab0ec7}.

\bibitem[{Mei} et~al.(2007){Mei}, {Blakeslee}, {C{\^o}t{\'e}}, {Tonry}, {West},
  {Ferrarese}, {Jord{\'a}n}, {Peng}, {Anthony}, and
  {Merritt}]{2007ApJ...655..144M}
S.~{Mei}, J.~P. {Blakeslee}, P.~{C{\^o}t{\'e}}, et~al.
\newblock {The ACS Virgo Cluster Survey. XIII. SBF Distance Catalog and the
  Three-dimensional Structure of the Virgo Cluster}.
\newblock \emph{\apj}, 655:\penalty0 144--162, January 2007.
\newblock \doi{10.1086/509598}.

\bibitem[{Beilicke} et~al.(2005){Beilicke}, {Cornils}, {Heinzelmann}, {Raue},
  {Ripken}, {Benbow}, {Horns}, {Tluczykont}, and {H.~E.~S.~S.
  Collaboration}]{2005tsra.conf..650B}
M.~{Beilicke}, R.~{Cornils}, G.~{Heinzelmann}, et~al.
\newblock {Observation of the Giant Radio Galaxy M87 at TeV Energies with
  H.E.S.S.}
\newblock In \emph{22nd Texas Symposium on Relativistic Astrophysics}, pages
  650--653, Jan 2005.

\bibitem[{Aleksi{\'c}} et~al.(2012{\natexlab{a}}){Aleksi{\'c}}, {Alvarez},
  {Antonelli}, {Antoranz}, {Asensio}, {Backes}, {Barrio}, {Bastieri}, {Becerra
  Gonz{\'a}lez}, and {Bednarek}]{2012A&A...544A..96A}
J.~{Aleksi{\'c}}, E.~A. {Alvarez}, L.~A. {Antonelli}, et~al.
\newblock {MAGIC observations of the giant radio galaxy M 87 in a low-emission
  state between 2005 and 2007}.
\newblock \emph{\aap}, 544:\penalty0 A96, Aug 2012{\natexlab{a}}.
\newblock \doi{10.1051/0004-6361/201117827}.

\bibitem[{Acciari} et~al.(2008){Acciari}, {Beilicke}, {Blaylock}, {Bradbury},
  {Buckley}, {Bugaev}, {Butt}, {Celik}, {Cesarini}, {Ciupik}, {Cogan}, {Colin},
  {Cui}, {Daniel}, {Duke}, {Ergin}, {Falcone}, {Fegan}, {Finley}, {Finnegan},
  {Fortin}, {Fortson}, {Gibbs}, {Gillanders}, {Grube}, {Guenette}, {Gyuk},
  {Hanna}, {Hays}, {Holder}, {Horan}, {Hughes}, {Hui}, {Humensky}, {Imran},
  {Kaaret}, {Kertzman}, {Kieda}, {Kildea}, {Konopelko}, {Krawczynski},
  {Krennrich}, {Lang}, {LeBohec}, {Lee}, {Maier}, {McCann}, {McCutcheon},
  {Millis}, {Moriarty}, {Mukherjee}, {Nagai}, {Ong}, {Pandel}, {Perkins},
  {Pohl}, {Quinn}, {Ragan}, {Reynolds}, {Rose}, {Schroedter}, {Sembroski},
  {Smith}, {Steele}, {Swordy}, {Syson}, {Toner}, {Valcarcel}, {Vassiliev},
  {Wakely}, {Ward}, {Weekes}, {Weinstein}, {White}, {Williams}, {Wissel},
  {Wood}, and {Zitzer}]{2008ApJ...679..397A}
V.~A. {Acciari}, M.~{Beilicke}, G.~{Blaylock}, et~al.
\newblock {Observation of Gamma-Ray Emission from the Galaxy M87 above 250 GeV
  with VERITAS}.
\newblock \emph{\apj}, 679:\penalty0 397--403, May 2008.
\newblock \doi{10.1086/587458}.

\bibitem[{Beilicke} and {VERITAS Collaboration}(2012)]{2012AIPC.1505..586B}
M.~{Beilicke} and {VERITAS Collaboration}.
\newblock {VERITAS observations of M87 in 2011/2012}.
\newblock In F.~A. {Aharonian}, W.~{Hofmann}, and F.~M. {Rieger}, editors,
  \emph{American Institute of Physics Conference Series}, volume 1505 of
  \emph{American Institute of Physics Conference Series}, pages 586--589,
  December 2012.
\newblock \doi{10.1063/1.4772328}.

\bibitem[{Aharonian} et~al.(2006){Aharonian}, {Akhperjanian}, {Bazer-Bachi},
  {Beilicke}, {Benbow}, {Berge}, {Bernl{\"o}hr}, {Boisson}, {Bolz}, {Borrel},
  {Braun}, {Brown}, {B{\"u}hler}, {B{\"u}sching}, {Carrigan}, {Chadwick},
  {Chounet}, {Coignet}, {Cornils}, {Costamante}, {Degrange}, {Dickinson},
  {Djannati-Ata{\"i}}, {Drury}, {Dubus}, {Egberts}, {Emmanoulopoulos},
  {Espigat}, {Feinstein}, {Ferrero}, {Fiasson}, {Fontaine}, {Funk}, {Funk},
  {F{\"u}{\ss}ling}, {Gallant}, {Giebels}, {Glicenstein}, {Goret},
  {Hadjichristidis}, {Hauser}, {Hauser}, {Heinzelmann}, {Henri}, {Hermann},
  {Hinton}, {Hoffmann}, {Hofmann}, {Holleran}, {Hoppe}, {Horns},
  {Jacholkowska}, {de Jager}, {Kendziorra}, {Kerschhaggl}, {Kh{\'e}lifi},
  {Komin}, {Konopelko}, {Kosack}, {Lamanna}, {Latham}, {Le Gallou},
  {Lemi{\`e}re}, {Lemoine-Goumard}, {Lenain}, {Lohse}, {Martin},
  {Martineau-Huynh}, {Marcowith}, {Masterson}, {Maurin}, {McComb}, {Moulin},
  {de Naurois}, {Nedbal}, {Nolan}, {Noutsos}, {Orford}, {Osborne}, {Ouchrif},
  {Panter}, {Pelletier}, {Pita}, {P{\"u}hlhofer}, {Punch}, {Ranchon},
  {Raubenheimer}, {Raue}, {Rayner}, {Reimer}, {Ripken}, {Rob}, {Rolland},
  {Rosier-Lees}, {Rowell}, {Sahakian}, {Santangelo}, {Saug{\'e}}, {Schlenker},
  {Schlickeiser}, {Schr{\"o}der}, {Schwanke}, {Schwarzburg}, {Schwemmer},
  {Shalchi}, {Sol}, {Spangler}, {Spanier}, {Steenkamp}, {Stegmann}, {Superina},
  {Tam}, {Tavernet}, {Terrier}, {Tluczykont}, {van Eldik}, {Vasileiadis},
  {Venter}, {Vialle}, {Vincent}, {V{\"o}lk}, {Wagner}, and
  {Ward}]{2006Sci...314.1424A}
F.~{Aharonian}, A.~G. {Akhperjanian}, A.~R. {Bazer-Bachi}, et~al.
\newblock {Fast Variability of Tera-Electron Volt {$\gamma$} Rays from the
  Radio Galaxy M87}.
\newblock \emph{Science}, 314:\penalty0 1424--1427, December 2006.
\newblock \doi{10.1126/science.1134408}.

\bibitem[{Albert} et~al.(2008){Albert}, {Aliu}, {Anderhub}, {Antonelli},
  {Antoranz}, {Backes}, {Baixeras}, {Barrio}, {Bartko}, {Bastieri}, {Becker},
  {Bednarek}, {Berger}, {Bernardini}, {Bigongiari}, {Biland}, {Bock},
  {Bonnoli}, {Bordas}, {Bosch-Ramon}, {Bretz}, {Britvitch}, {Camara},
  {Carmona}, {Chilingarian}, {Commichau}, {Contreras}, {Cortina}, {Costado},
  {Covino}, {Curtef}, {Dazzi}, {De Angelis}, {De Cea del Pozo}, {de los Reyes},
  {De Lotto}, {De Maria}, {De Sabata}, {Delgado Mendez}, {Dominguez}, {Dorner},
  {Doro}, {Errando}, {Fagiolini}, {Ferenc}, {Fern{\'a}ndez}, {Firpo},
  {Fonseca}, {Font}, {Galante}, {Garc{\'{\i}}a L{\'o}pez}, {Garczarczyk},
  {Gaug}, {Goebel}, {Hayashida}, {Herrero}, {H{\"o}hne}, {Hose}, {Hsu},
  {Huber}, {Jogler}, {Kranich}, {La Barbera}, {Laille}, {Leonardo}, {Lindfors},
  {Lombardi}, {Longo}, {L{\'o}pez}, {Lorenz}, {Majumdar}, {Maneva},
  {Mankuzhiyil}, {Mannheim}, {Maraschi}, {Mariotti}, {Mart{\'{\i}}nez},
  {Mazin}, {Meucci}, {Meyer}, {Miranda}, {Mirzoyan}, {Mizobuchi}, {Moles},
  {Moralejo}, {Nieto}, {Nilsson}, {Ninkovic}, {Otte}, {Oya}, {Panniello},
  {Paoletti}, {Paredes}, {Pasanen}, {Pascoli}, {Pauss}, {Pegna},
  {Perez-Torres}, {Persic}, {Peruzzo}, {Piccioli}, {Prada}, {Prandini},
  {Puchades}, {Raymers}, {Rhode}, {Rib{\'o}}, {Rico}, {Rissi}, {Robert},
  {R{\"u}gamer}, {Saggion}, {Saito}, {Salvati}, {Sanchez-Conde}, {Sartori},
  {Satalecka}, {Scalzotto}, {Scapin}, {Schweizer}, {Shayduk}, {Shinozaki},
  {Shore}, {Sidro}, {Sierpowska-Bartosik}, {Sillanp{\"a}{\"a}}, {Sobczynska},
  {Spanier}, {Stamerra}, {Stark}, {Takalo}, {Tavecchio}, {Temnikov}, {Tescaro},
  {Teshima}, {Tluczykont}, {Torres}, {Turini}, {Vankov}, {Venturini}, {Vitale},
  {Wagner}, {Wittek}, {Zabalza}, {Zandanel}, {Zanin}, and
  {Zapatero}]{2008ApJ...685L..23A}
J.~{Albert}, E.~{Aliu}, H.~{Anderhub}, et~al.
\newblock {Very High Energy Gamma-Ray Observations of Strong Flaring Activity
  in M87 in 2008 February}.
\newblock \emph{\apjl}, 685:\penalty0 L23, September 2008.
\newblock \doi{10.1086/592348}.

\bibitem[{Acciari} et~al.(2010){Acciari}, {Aliu}, {Arlen}, {Aune}, {Beilicke},
  {Benbow}, {Boltuch}, {Bradbury}, {Buckley}, {Bugaev}, {Byrum}, {Cannon},
  {Cesarini}, {Chow}, {Ciupik}, {Cogan}, {Cui}, {Dickherber}, {Duke}, {Finley},
  {Finnegan}, {Fortin}, {Fortson}, {Furniss}, {Galante}, {Gall}, {Gillanders},
  {Godambe}, {Grube}, {Guenette}, {Gyuk}, {Hanna}, {Holder}, {Hui}, {Humensky},
  {Imran}, {Kaaret}, {Karlsson}, {Kertzman}, {Kieda}, {Konopelko},
  {Krawczynski}, {Krennrich}, {Lang}, {LeBohec}, {Maier}, {McArthur}, {McCann},
  {McCutcheon}, {Millis}, {Moriarty}, {Ong}, {Otte}, {Pandel}, {Perkins},
  {Pichel}, {Pohl}, {Quinn}, {Ragan}, {Reyes}, {Reynolds}, {Roache}, {Rose},
  {Rovero}, {Schroedter}, {Sembroski}, {Senturk}, {Smith}, {Steele}, {Swordy},
  {Theiling}, {Thibadeau}, {Varlotta}, {Vincent}, {Wagner}, {Wakely}, {Ward},
  {Weekes}, {Weinstein}, {Weisgarber}, {Williams}, {Wissel}, {Wood}, {Zitzer},
  {Harris}, and {Massaro}]{2010ApJ...716..819A}
V.~A. {Acciari}, E.~{Aliu}, T.~{Arlen}, et~al.
\newblock {Veritas 2008-2009 Monitoring of the Variable Gamma-ray Source M 87}.
\newblock \emph{\apj}, 716:\penalty0 819--824, June 2010.
\newblock \doi{10.1088/0004-637X/716/1/819}.

\bibitem[{Aliu} et~al.(2012){Aliu}, {Arlen}, {Aune}, {Beilicke}, {Benbow},
  {Bouvier}, {Bradbury}, {Buckley}, {Bugaev}, {Byrum}, {Cannon}, {Cesarini},
  {Ciupik}, {Collins-Hughes}, {Connolly}, {Cui}, {Dickherber}, {Duke},
  {Errando}, {Falcone}, {Finley}, {Finnegan}, {Fortson}, {Furniss}, {Galante},
  {Gall}, {Godambe}, {Griffin}, {Grube}, {Guenette}, {Gyuk}, {Hanna}, {Holder},
  {Huan}, {Hughes}, {Hui}, {Humensky}, {Imran}, {Kaaret}, {Karlsson},
  {Kertzman}, {Kieda}, {Krawczynski}, {Krennrich}, {Lang}, {LeBohec},
  {Madhavan}, {Maier}, {Majumdar}, {McArthur}, {McCann}, {Moriarty},
  {Mukherjee}, {Nu{\~n}ez}, {Ong}, {Orr}, {Otte}, {Park}, {Perkins}, {Pichel},
  {Pohl}, {Prokoph}, {Quinn}, {Ragan}, {Reyes}, {Reynolds}, {Roache}, {Rose},
  {Ruppel}, {Saxon}, {Schroedter}, {Sembroski}, {{\c S}ent{\"u}rk}, {Skole},
  {Staszak}, {Te{\v s}i{\'c}}, {Theiling}, {Thibadeau}, {Tsurusaki}, {Tyler},
  {Varlotta}, {Vassiliev}, {Vincent}, {Vivier}, {Wakely}, {Ward}, {Weekes},
  {Weinstein}, {Weisgarber}, {Williams}, and {Zitzer}]{2012ApJ...746..141A}
E.~{Aliu}, T.~{Arlen}, T.~{Aune}, et~al.
\newblock {VERITAS Observations of Day-scale Flaring of M 87 in 2010 April}.
\newblock \emph{\apj}, 746:\penalty0 141, February 2012.
\newblock \doi{10.1088/0004-637X/746/2/141}.

\bibitem[{Strauss} et~al.(1992){Strauss}, {Huchra}, {Davis}, {Yahil}, {Fisher},
  and {Tonry}]{1992ApJS...83...29S}
M.~A. {Strauss}, J.~P. {Huchra}, M.~{Davis}, et~al.
\newblock {A redshift survey of IRAS galaxies. VII - The infrared and redshift
  data for the 1.936 Jansky sample}.
\newblock \emph{\apjs}, 83:\penalty0 29--63, November 1992.
\newblock \doi{10.1086/191730}.

\bibitem[{Wilman} et~al.(2005){Wilman}, {Edge}, and
  {Johnstone}]{2005MNRAS.359..755W}
R.~J. {Wilman}, A.~C. {Edge}, and R.~M. {Johnstone}.
\newblock {The nature of the molecular gas system in the core of NGC 1275}.
\newblock \emph{\mnras}, 359:\penalty0 755--764, May 2005.
\newblock \doi{10.1111/j.1365-2966.2005.08956.x}.

\bibitem[{Aleksi{\'c}} et~al.(2012{\natexlab{b}})]{2012A&A...539L...2A}
J.~{Aleksi{\'c}} et~al.
\newblock {Detection of very-high energy {$\gamma$}-ray emission from
  <ASTROBJ>NGC 1275</ASTROBJ> by the MAGIC telescopes}.
\newblock \emph{\aap}, 539:\penalty0 L2, March 2012{\natexlab{b}}.
\newblock \doi{10.1051/0004-6361/201118668}.

\bibitem[{MAGIC Collaboration} et~al.(2018){MAGIC Collaboration}, {Ansoldi},
  et~al.]{2018A&A...617A..91M}
{MAGIC Collaboration}, S.~{Ansoldi}, et~al.
\newblock {Gamma-ray flaring activity of NGC1275 in 2016-2017 measured by
  MAGIC}.
\newblock \emph{\aap}, 617:\penalty0 A91, September 2018.
\newblock \doi{10.1051/0004-6361/201832895}.

\bibitem[{Struble} and {Rood}(1999)]{1999ApJS..125...35S}
M.~F. {Struble} and H.~J. {Rood}.
\newblock {A Compilation of Redshifts and Velocity Dispersions for ACO
  Clusters}.
\newblock \emph{\apjs}, 125:\penalty0 35--71, November 1999.
\newblock \doi{10.1086/313274}.

\bibitem[{de Ruiter} et~al.(2015){de Ruiter}, {Parma}, {Fanti}, and
  {Fanti}]{2015A&A...581A..33D}
H.~R. {de Ruiter}, P.~{Parma}, R.~{Fanti}, and C.~{Fanti}.
\newblock {Far-UV to mid-IR properties of nearby radio galaxies}.
\newblock \emph{\aap}, 581:\penalty0 A33, September 2015.
\newblock \doi{10.1051/0004-6361/201424079}.

\bibitem[{Meyer} et~al.(2015){Meyer}, {Georganopoulos}, {Sparks}, {Perlman},
  {van der Marel}, {Anderson}, {Sohn}, {Biretta}, {Norman}, and
  {Chiaberge}]{2015Natur.521..495M}
E.~T. {Meyer}, M.~{Georganopoulos}, W.~B. {Sparks}, et~al.
\newblock {A kiloparsec-scale internal shock collision in the jet of a nearby
  radio galaxy}.
\newblock \emph{\nat}, 521:\penalty0 495--497, May 2015.
\newblock \doi{10.1038/nature14481}.

\bibitem[{Mukherjee}(2018)]{2018ATel11436....1M}
R.~{Mukherjee}.
\newblock {VERITAS discovery of VHE emission from the FRI radio galaxy 3C 264}.
\newblock \emph{The Astronomer's Telegram}, 11436, March 2018.

\bibitem[Rieger and Levinson(2018)]{2018arXiv181005409R}
Frank~M. Rieger and Amir Levinson.
\newblock Radio galaxies at vhe energies.
\newblock \emph{Galaxies}, 6\penalty0 (4), 2018.
\newblock ISSN 2075-4434.
\newblock \doi{10.3390/galaxies6040116}.
\newblock URL \url{https://www.mdpi.com/2075-4434/6/4/116}.

\bibitem[{Abeysekara} et~al.(2017{\natexlab{a}}){Abeysekara}, {Albert},
  {Alfaro}, {Alvarez}, {{\'A}lvarez}, {Arceo}, {Arteaga-Vel{\'a}zquez}, {Ayala
  Solares}, {Barber}, {Bautista-Elivar}, {Becerril}, {Belmont-Moreno},
  {BenZvi}, {Berley}, {Braun}, {Brisbois}, {Caballero-Mora}, {Capistr{\'a}n},
  {Carrami{\~n}ana}, {Casanova}, {Castillo}, {Cotti}, {Cotzomi}, {Couti{\~n}o
  de Le{\'o}n}, {de la Fuente}, {De Le{\'o}n}, {DeYoung}, {Dingus},
  {DuVernois}, {D{\'{\i}}az-V{\'e}lez}, {Ellsworth}, {Fiorino}, {Fraija},
  {Garc{\'{\i}}a-Gonz{\'a}lez}, {Gerhardt}, {Gonz{\'a}lez Mun{\"o}z},
  {Gonz{\'a}lez}, {Goodman}, {Hampel-Arias}, {Harding}, {Hernandez},
  {Hernandez-Almada}, {Hinton}, {Hui}, {H{\"u}ntemeyer}, {Iriarte},
  {Jardin-Blicq}, {Joshi}, {Kaufmann}, {Kieda}, {Lara}, {Lauer}, {Lee},
  {Lennarz}, {Le{\'o}n Vargas}, {Linnemann}, {Longinotti}, {Raya},
  {Luna-Garc{\'{\i}}a}, {L{\'o}pez-Coto}, {Malone}, {Marinelli}, {Martinez},
  {Martinez-Castellanos}, {Mart{\'{\i}}nez-Castro}, {Mart{\'{\i}}nez-Huerta},
  {Matthews}, {Miranda-Romagnoli}, {Moreno}, {Mostaf{\'a}}, {Nellen},
  {Newbold}, {Nisa}, {Noriega-Papaqui}, {Pelayo}, {Pretz},
  {P{\'e}rez-P{\'e}rez}, {Ren}, {Rho}, {Rivi{\`e}re}, {Rosa-Gonz{\'a}lez},
  {Rosenberg}, {Ruiz-Velasco}, {Salazar}, {Salesa Greus}, {Sandoval},
  {Schneider}, {Schoorlemmer}, {Sinnis}, {Smith}, {Springer}, {Surajbali},
  {Taboada}, {Tibolla}, {Tollefson}, {Torres}, {Ukwatta}, {Villase{\~n}or},
  {Weisgarber}, {Westerhoff}, {Wisher}, {Wood}, {Yapici}, {Yodh}, {Younk},
  {Zepeda}, and {Zhou}]{2017ApJ...843...39A}
A.~U. {Abeysekara}, A.~{Albert}, R.~{Alfaro}, et~al.
\newblock {Observation of the Crab Nebula with the HAWC Gamma-Ray Observatory}.
\newblock \emph{\apj}, 843:\penalty0 39, July 2017{\natexlab{a}}.
\newblock \doi{10.3847/1538-4357/aa7555}.

\bibitem[{Younk} et~al.(2015){Younk}, {Lauer}, {Vianello}, {Harding}, {Ayala
  Solares}, {Zhou}, and {Michelle Hui for the HAWC
  Collaboration}]{2015arXiv150807479Y}
Patrick~W. {Younk}, Robert~J. {Lauer}, Giacomo {Vianello}, et~al.
\newblock {A high-level analysis framework for HAWC}.
\newblock \emph{arXiv e-prints}, art. arXiv:1508.07479, Aug 2015.

\bibitem[{Abeysekara} et~al.(2017{\natexlab{b}}){Abeysekara}, {Albert},
  {Alfaro}, {Alvarez}, {{\'A}lvarez}, {Arceo}, {Arteaga-Vel{\'a}zquez}, {Avila
  Rojas}, {Ayala Solares}, {Barber}, {Bautista-Elivar}, {Becerra Gonzalez},
  {Becerril}, {Belmont-Moreno}, {BenZvi}, {Bernal}, {Braun}, {Brisbois},
  {Caballero-Mora}, {Capistr{\'a}n}, {Carrami{\~n}ana}, {Casanova}, {Castillo},
  {Cotti}, {Cotzomi}, {Couti{\~n}o de Le{\'o}n}, {De Le{\'o}n}, {De la Fuente},
  {Diaz Hernandez}, {Dingus}, {DuVernois}, {D{\'{\i}}az-V{\'e}lez},
  {Ellsworth}, {Engel}, {Fiorino}, {Fraija}, {Garc{\'{\i}}a-Gonz{\'a}lez},
  {Garfias}, {Gerhardt}, {Gonz{\'a}lez Mu{\~n}oz}, {Gonz{\'a}lez}, {Goodman},
  {Hampel-Arias}, {Harding}, {Hernandez}, {Hernandez-Almada}, {Hona}, {Hui},
  {H{\"u}ntemeyer}, {Iriarte}, {Jardin-Blicq}, {Joshi}, {Kaufmann}, {Kieda},
  {Lara}, {Lauer}, {Lee}, {Lennarz}, {Le{\'o}n Vargas}, {Linnemann},
  {Longinotti}, {Raya}, {Luna-Garc{\'{\i}}a}, {L{\'o}pez-Coto}, {Malone},
  {Marinelli}, {Martinez}, {Martinez-Castellanos}, {Mart{\'{\i}}nez-Castro},
  {Matthews}, {Miranda-Romagnoli}, {Moreno}, {Mostaf{\'a}}, {Nellen},
  {Newbold}, {Nisa}, {Noriega-Papaqui}, {Pretz}, {P{\'e}rez-P{\'e}rez}, {Ren},
  {Rho}, {Rivi{\`e}re}, {Rosa-Gonz{\'a}lez}, {Rosenberg}, {Ruiz-Velasco},
  {Salesa Greus}, {Sandoval}, {Schneider}, {Schoorlemmer}, {Sinnis}, {Smith},
  {Springer}, {Surajbali}, {Taboada}, {Tibolla}, {Tollefson}, {Torres},
  {Ukwatta}, {Vianello}, {Weisgarber}, {Westerhoff}, {Wisher}, {Wood},
  {Yapici}, {Younk}, {Zepeda}, and {Zhou}]{2017ApJ...841..100A}
A.~U. {Abeysekara}, A.~{Albert}, R.~{Alfaro}, et~al.
\newblock {Daily Monitoring of TeV Gamma-Ray Emission from Mrk 421, Mrk 501,
  and the Crab Nebula with HAWC}.
\newblock \emph{\apj}, 841:\penalty0 100, June 2017{\natexlab{b}}.
\newblock \doi{10.3847/1538-4357/aa729e}.

\bibitem[{Garc\'ia-Gonz\'alez} et~al.(){Garc\'ia-Gonz\'alez}, {Fraija}, and
  {Gonz\'alez for the HAWC Collaboration}]{JA_ICRC}
J.~A. {Garc\'ia-Gonz\'alez}, N.~{Fraija}, and M.M. {Gonz\'alez for the HAWC
  Collaboration}.
\newblock {Gamma-ray light curves for the BL Lac Mrk 421 using HAWC data
  derived with a new approach.}

\end{thebibliography}

\end{document}